\documentclass[final,authoryear,1p,times]{elsarticle}

\usepackage{graphicx}

\usepackage{rotating}

\usepackage{amssymb}
\usepackage{amsthm}

\journal{New Astronomy}

\begin{document}

\begin{frontmatter}



\title{Ion-cyclotron waves in Solar Coronal Hole}


\author[label1,label2]{{S. Do\u{g}an}\corref{cor1}}
\ead{suzan.dogan@mail.ege.edu.tr} \cortext[cor1]{Corresponding
author}
 \address[label1]{University of Ege, Faculty of Science, Department of Astronomy and Space Sciences, Bornova, 35100, \.Izmir, Turkey}
  \address[label2]{Theoretical Astrophysics Group, University of
Leicester, Leicester, LE1 7RH, UK}

\author[label1]{E. R. Pek\"{u}nl\"{u}}

\address{}

\begin{abstract}

We investigate the effect of the Plume/Interplume Lane (PIPL)
structure of the solar Polar Coronal Hole (PCH) on the propagation
characteristics of ion-cyclotron waves (ICW). The gradients of
physical parameters determined by SOHO and TRACE satellites both
parallel and perpendicular to the magnetic field are considered with
the aim of determining how the efficiency of the ICR process varies
along the PIPL structure of PCH. We construct a model based on the
kinetic theory by using quasi-linear approximation. We solve the
Vlasov equation for O VI ions and obtain the dispersion relation of
ICW. The resonance process in the interplume lanes is much more
effective than in the plumes, agreeing with the observations which
show the source of fast solar wind is interplume lanes. The solution
of the Vlasov equation in PIPL structure of PCH, the physical
parameters of which display gradients along and perpendicular
direction to the external magnetic field, is thus obtained in a more
general form than the previous investigations.
\end{abstract}

\begin{keyword}
Sun: corona - solar wind - plasmas - waves - acceleration of particles


\end{keyword}

\end{frontmatter}


\section{Introduction}
 The motivation for the present investigation is given by UVCS/SOHO measurements which showed the collisionless
nature of the solar North Polar Coronal Hole (NPCH) (e.g. Cranmer et
al. 1999; Kohl et al. 1999). What was seen in the Mg X and O VI line
measurements was that the plasma in NPCH was collisionless beyond
the 1.75 - 2.1 $\rm R_{\odot}$ (Doyle et al. 1999). Kumar et al.
(2006) took the dissipative terms like resistivity and viscosity
into account and considered the parallel (to the magnetic field)
heat conduction due to Coulomb collisions in order to investigate
the coronal heating by MHD waves. All the efforts have been spent to
account for the heating of the different parts of corona by
wave-particle interactions.

Tu and Marsch (1997) developed a two-fluid MHD model of the solar
corona and wind. In their model they investigated the effects of
Alfven waves, within the 1 Hz-1 kHz frequency range, on the heating
and acceleration. They showed that their model can fit the observed
density profiles of polar coronal hole and polar plume at 2.4 $\rm
R_{\odot}$ and 3.2 $\rm R_{\odot}$, respectively. They also obtained
the solar wind speeds at 63 $\rm R_{\odot}$.

Marsch and Tu (1997) investigated the heating and acceleration of
the solar corona and the solar wind by considering Alfven waves that
are assumed to be produced by small scale reconnections in the
chromospheric network. Their two-fluid model predicted that the
Alfven waves within the 200 Hz - 790 Hz frequency range damp inside
1.5 $\rm R_{\odot}$ and thus heat the corona and accelerate the
solar wind near the Sun.

Besides, NPCH is also structured in radial direction by so-called
plume and interplume lanes (PIPL) (see, Figure 1 in Wilhelm et al.
1998). PIPL structure is taken into account by several authors in
their investigation (e.g., Ofman et al. 2000). Upper chromosphere
and the corona show anisotropic viscosity, resistivity and thermal
conductivity. Ruderman et al. (2000) considered the damping of the
slow surface waves propagating along the magnetic field in such a
medium.

Lie-Svendsen et al. (2002) studied the solar wind in expanding
coronal holes wherein the heat flux densities are along the magnetic
field. By the accumulation of data on NPCH, it has been clear that
the coronal hole heating by MHD waves should be considered in the
\emph{nonclassical} context, i.e., Coulomb collisions and Maxwell
distribution of plasma species cannot and should not be assumed
(Williams 1997).

It is a well-known fact that in such a medium like NPCH where the
particle number density as well as the effective temperature of the
plasma species have gradients both in the radial and the
perpendicular direction to the line of sight (\emph{los}) and the
magnetic field has a gradient in radial direction, the wave
propagation characteristics display novelties (e.g., Joarder et al.
1997). If the typical length scale of the wave is comparable to the
length scales of the physical parameters of the medium MHD waves
tend to refract towards resonance and go through collisionless
damping (Melrose \& McPhedron 1991).

Nakariakov et al. (1998), in dealing with the coronal heating
processes of Sun and late type stars, considered nonlinear wave
coupling in a MHD context. They retained the nonlinear terms in the
x and z-components of the linearized momentum equation. They claimed
that the nonlinear terms in their equations are to be held
responsible for the generation of fast modes. They also cite their
previous studies wherein they showed that the cause of the nonlinear
generation of these fast waves are the longitudinal and transversal
gradients in the magnetic pressure. Nakariakov (2006) considered the
effects of transverse structuring in plumes and drew attention to
the effect of different spatial scales of transverse and
longitudinal inhomogeneities on the wave dynamics. We are agreed
with Nakariakov and propose to take into account the gradients of
physical parameters both in plume and interplume lanes.

Vocks \& Marsch (2001) treated the wave-particle interaction within
the framework of quasilinear theory by solving the Vlasov equation
numerically. They considered the heating process by a wave-particle
interaction in a coronal funnel and in the lower corona up to a
height of $\rm 0.57R_{\odot}$. Their Figure 1 shows that $T_{\bot}$
for $\rm{O}^{5+}$ reaches about $3\times10^{7}$ K at the upper part
of the computational domain. They assumed dispersionless waves and
put off the consideration of dispersive effects for a future
investigation.

In the present investigation, we also solve the Vlasov equation in
2D one of which covers the distance range of 1.5R - 3.5R, where R is
the dimensionless radial distance, i.e., $\rm R=r/R_{\odot}$ and the
other is the x-direction perpendicular both to the radial direction
and the \emph{los} and the details of x is given in Subsection 2.1.
Wave propagation characteristics in NPCH was investigated by
Pek\"{u}nl\"{u} et al. (2004). They did not take the PIPL structure
of the medium into account. To the best knowledge of the authors of
the present investigation the effect of PIPL structure of NPCH on
the wave propagation characteristics has not been taken into account
hitherto. This is our aim in this study. We should mention in
passing that in our investigation, we take the observational
parameters of NPCH into consideration. Nevertheless, the same
conclusions we draw apply also all the coronal holes be it as it may
appear in the north or south. Since our model should be valid for
polar coronal hole (PCH) in general, we will use the term PCH for
our investigation.

Cranmer et al. (2008) list the three primary observables as the
absolute intensity of the O VI $\lambda1032$ line, the line width
$v_{1/2}$ and the intensity ratio R of the  $\lambda1032$ to the
$\lambda1037$, which depend on the four or \textquotedblleft
unknown\textquotedblright quantities along the \emph{los}
distribution, i.e., ion fraction $n_{\rm{OVI}}/n_{\rm{e}}$, the O VI
bulk outflow speed along the magnetic field, and the parallel and
perpendicular $\rm{O}^{5+}$ kinetic temperatures. They refrain from
making a definitive separation between $T_{\rm{i}}$ the ionization
temperature and the nonthermal part of the effective temperature
(see Eq. 1). They also draw attention to the necessity for new
observations in order to make further progress and not to make
arbitrary assumptions about, e.g., the ion outflow speed or the ion
temperature. Kohl et al. (2006) showed that the plasma properties in
NPCH remain reasonably constant about a year or two around solar
minimum (1996-1997). This constancy of plasma parameters enables us
to express them as a function of R and x.

Isenberg \& Vosquez (2009) considered the cyclotron-resonant Fermi
heating mechanism by taking into account the effects of gravity,
charge-separation electric field, and mirror force in a radially
expanding flux tube. They found that a small fraction of nonlinearly
generated resonant wave power can provide the observed energization
for O VI ions.

Devlen $\&$ Pek{\"u}nl{\"u} (2010) investigated the effects of the O
VI temperature and number density gradients in the parallel and
perpendicular direction to the magnetic field in the MHD context.
Their results show that the perpendicular (to the external magnetic
field) heat conduction introduces novelty to the wave propagation
characteristics.

The plan of the paper is as follows: In Section 2 we review the
plasma properties of the NPCH revealed by SOHO and TRACE satellites.
In Section 3, we assume that the coronal hole plasma is electrically
\emph{quasineutral}, that is, $N_{\rm{e}}\approx N_{\rm{p}}\approx
N$ where subscripts stand for electron and proton, respectively
(Marsch 1999; Endeve \& Leer 2001; Voitenko \& Goosens 2002). Then,
bearing in mind another observational fact that O VI ions are
preferentially heated, we'll obtain the dispersion relation of the
ion cyclotron waves (ICW) in PIPL structure of the PCH by solving
the Vlasov equation and present the summary and conclusion in
Section 4. We use CGS units in this study.

\section{NPCH Plasma Properties}
\subsection{O VI Ion Temperatures}
Since the NPCH is collisionless and a typical low - $\beta$ plasma,
we expect that it shows temperature anisotropy. Indeed, O VI  1032
{\AA} line intensities vary in the radial direction as well as in
the direction both perpendicular to the radial and the \emph{los}
ones (Kohl et al., 1997a). Their Figure 16 shows that O VI line
widths obtained from the darker interplume lanes are wider than the
ones coming from the brighter plumes. SUMER observations also
revealed that the line widths from interplume lanes are wider than
those of plumes (Wilhelm et al. 2000; Banerjee et al. 2000, Banerjee
et al. 2009a). The effective temperature of the O VI ions at 3.0 R
is given as $T_{\bot {\rm{eff}}}\sim10^{8}$K (Cranmer et al. 1999).
Antonucci et al. (2000) give the effective temperature profile of O
VI ions in the 1.5 - 3.0 R. The relation between the ion temperature
and the effective temperature is,
 \begin{equation}
T_{\rm{eff}} =(m_{\rm i}/2k_{\rm B})v_{\rm 1/e}^{2}=T_{\rm{i}}+({m}_{\rm i}/2{k}_{\rm B})\xi^{2}
\end{equation}
where $\rm{T}_{i}$ is the ion temperature; $\rm{m}_{i}$ is the ion
mass; $\rm{k}_{B}$ is the Boltzmann constant; $v_{1/e}$ is the
\emph{los} speed; $\xi$ is the most probable speed of an isotropic,
Gaussian-distributed, turbulent velocity field (Wilhelm et al.
1998). The relation between $\xi$ and the wave amplitude is
$\xi^{2}=(1/2)\langle\delta v^{2}\rangle$ (Esser et al. 1999). Doyle
et al. (1999) report that for Alfven-like waves \textquotedblleft
factor 2 accounts for the polarization and direction of propagation
of a wave relative to the \emph{los}\textquotedblright.

We follow the coup and do not make any assumptions about the ion
temperature and the nonthermal part of the effective temperature but
only stick to the observational fact that there is no temporal
variation of the effective temperature within several time scales of
ICW's propagation in NPCH.

Banerjee et al. (2000) used the comprehensive and self-consistent
empirical model of Cranmer et al. (1999) and derived a best-fit
function for the O VI line width valid in 1.5-3.5 R (see their Eq.
3). We converted their Eq. 3 into the effective temperature as
below,
\begin{equation}
T_{\rm{eff}}(R)=4.02 \times 10^{7}R^{2}+1.25 \times 10^{7}R-9.76 \times 10^{7} \,\,\, \rm K.
\end{equation}
Since there is no data about the effective temperature of NPCH
plasma and/or O VI ions along the \emph{los} we assumed that
\emph{los} temperature of O VI ions is constant. Esser et al. (1999)
showed that Mg X and O VI ion temperatures ($T_{\rm i}$) are much
greater than the proton temperature. They also compared the ion
temperatures of Mg X and O VI ions on various distances from the
solar surface. For instance, they found that at 1.6 and 1.75 $\rm
R_{\odot}$, O VI temperature is in the same range as $T_{\rm Mg X}$.
But at 1.9 and 2 $\rm R_{\odot}$ $T_{\rm O VI}$ was found to be much
higher than $T_{\rm Mg X}$. Ion temperatures ratio of these two
minor ions displayed a minimum value of 1.6 at 1.9 $\rm R_{\odot}$
and 1.32 at 2 $\rm R_{\odot}$ and the maximum value of 13 at 1.9
$\rm R_{\odot}$ and 6.6 at 2 $\rm R_{\odot}$. We used Fig. 2a in
Esser et al. (1999) in order to find polynomial form for the radial
profile $\delta v_{\rm Mg X}$. We tried hard to find a similar
radial profile for O VI in the literature, but in vain. Therefore we
adopt $\delta v_{\rm Mg X}$ as if it is $\delta v_{\rm O VI}$ and
derived the polynomial form of the nonthermal part of $T_{\rm eff}$
of O VI ions and designate it as $T_{\rm eff}^{\xi}$:

\begin{table}
\begin{center}
\caption{O VI Temperatures}
\begin{tabular}{c c c c c}
\hline
\hline
R & $T_{\rm eff}$  & $T_{\rm i}$  & $T_{\rm eff}^{\xi}$ & $T_{\rm eff}^{\xi}/T_{\rm i}$  \\
$(r/R_{\odot})$& ($/10^{7}$ K)  & ($/10^{7}$ K)  & ($/10^{7}$ K) &   \\

\hline
1.5 &   1.6 &   0.80    &   0.77    &   0.96    \\
1.6 &   3.0 &   2.1 &   0.9 &   0.43    \\
1.7 &   4.5 &   3.5 &   1.0 &   0.29    \\
1.8 &   6.1 &   5.0 &   1.1 &   0.23    \\
1.9 &   7.8 &   6.5 &   1.3 &   0.19    \\
2.0 &   9.5 &   8.2 &   1.4 &   0.17    \\
2.1 &   11.4    &   9.9 &   1.5 &   0.15    \\
2.2 &   13.3    &   11.7    &   1.6 &   0.13    \\
2.3 &   15.3    &   13.7    &   1.7 &   0.12    \\
2.4 &   17.4    &   15.7    &   1.8 &   0.11    \\
2.5 &   19.6    &   17.8    &   1.8 &   0.10    \\
2.6 &   21.9    &   20.0    &   1.9 &   0.10    \\
2.7 &   24.2    &   22.2    &   2.0 &   0.09    \\
2.8 &   26.7    &   24.6    &   2.1 &   0.08    \\
2.9 &   29.2    &   27.0    &   2.1 &   0.08    \\
3.0 &   31.8    &   29.6    &   2.2 &   0.07    \\
3.1 &   34.5    &   32.2    &   2.3 &   0.07    \\
3.2 &   37.2    &   34.9    &   2.3 &   0.07    \\
3.3 &   40.1    &   37.8    &   2.4 &   0.06    \\
3.4 &   43.0    &   40.7    &   2.4 &   0.06    \\
3.5 &   46.1    &   43.6    &   2.4 &   0.06    \\
\hline

\end{tabular}

\medskip
Effective temperature ($T_{\rm eff}$), ion temperature ($T_{\rm i}$)
and the non-thermal part of the effective temperature ($T_{\rm
eff}^{\xi}$) of O VI ions are listed as a function of radial
distance (R).
\end{center}
\end{table}

\begin{equation}
T_{\rm eff}^{\xi}=-2.48 \times 10^{6}R^{2}+2.07 \times 10^{7}R-1.78\times 10^{7}\,\,\, \rm K.
 \end{equation}

Using $\delta v_{\rm Mg X}$ profile to represent $\delta v_{\rm O
VI}$ in our calculations surely introduces some imprecision to our
results but it is less than, at most, and order of magnitude, so we
did not refrain from adopting $\delta v_{\rm Mg X}$ for $\delta
v_{\rm O VI}$. In Table 1 we present the variations of $T_{\rm
eff}$, $T_{\rm i}$, $T_{\rm eff}^{\xi}$ and $T_{\rm
eff}^{\xi}/T_{\rm i}$ with the radial distance. $T_{\rm eff}$ is
derived from the O VI line width (Banerjee et al. 2000; Cranmer et
al. 1999); $T_{\rm eff}^{\xi}$ is from Esser et al. (1999). By using
these values in Eq. (4) we obtained $T_{\rm i}$.

Devlen $\&$ Pek{\"u}nl{\"u} (2010) modeled the non-thermal part of
the effective temperature of O VI ions in two dimension in NPCH (see
Fig. 1). Wilhelm et al. (1998) report that the effective temperature
of O VI ions in the interplume lanes are about $30 \%$ higher than
that of the plumes. We derived the length scale of the effective
temperature gradient perpendicular to both the radial direction and
the \emph{los} by referring to the Figure 1 of Wilhelm et al.
(1998). They give the width of the PIPL structure of the NPCH at
1.03 R as $380''$. At Sun, $1''\thickapprox    \rm{715\, km}$. There
appears four interplume lanes and four plumes in Figure 1 of Wilhelm
et al. (1998). Although the widths of plumes and interplume lanes
are unequal, we made a rough estimation and divided $380''$  by
eight and assumed that they are equal in width. The average width of
these plumes and interplume lanes turns out to be 33962.5 km. We
take this value as the length scale $(\Lambda_{\rm x}^{T})$ of the
perpendicular temperature gradient at 1.03 R. Between R = 1.034 and
R = 1.32, the widths of the plumes are reported to increase by a
factor of 2 (Wilhelm et al. 1998). This observational fact enables
Devlen \& Pek{\"u}nl{\"u} (2010) to formulate the temperature
structure of NPCH in (\emph{R, x}) space as given by Eq. (4).
 \begin{figure}
\centering
\includegraphics[scale=0.35]{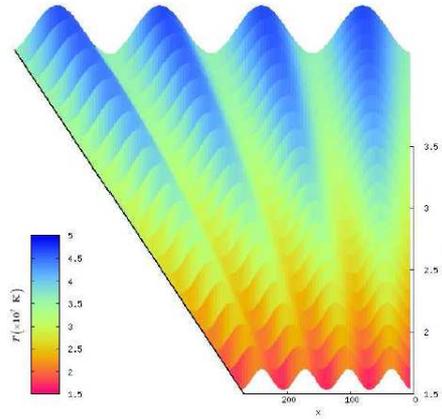}
\caption{Non-thermal part of the effective temperature of O VI ions
in NPCH as a function of R and x. The crests corresponds to the
interplume lanes and the troughs to the plumes. The width of the
PIPL varies with R. The abscissa x is in arcseconds as it is also in
Wilhelm et al. (1998). From Devlen \& Pek\"{u}nl\"{u} (2010).}
\end{figure}
\begin{equation}
T_{\rm{eff}}(R,x)= T_{\rm{eff}}(R)+ 0.3T_{\rm{eff}}(R)\sin^{2}(\frac {2\pi}{\lambda} x)
\end{equation}
where \emph{x} is the direction perpendicular both to R and
\emph{los} and $\lambda$ is the expansion rate of the widths of the
PIPL in NPCH in arc seconds, i.e., $\lambda= 92''.16 R$ (see Fig.
1). $\lambda$ also represents the wavelengths of the $\sin^{2}(2\pi
x/\lambda)$ function which is a function of R and x. The factor 0.3
comes from the difference between the plume and interplume lane
temperatures (30$\%$ max.). In that case, the effective temperature
gradient of O VI ions in two dimensions can be expressed as below
(Devlen $\&$ Pek{\"u}nl{\"u}, 2010):
\begin{equation}
\nabla T_{\rm eff}^{\xi}(R,x)=\frac{\partial T_{\rm eff}^{\xi}}{\partial R}\widehat{\textbf{R}}+\frac{\partial T_{\rm eff}^{\xi}}{\partial x}\widehat{\textbf{x}}
 \end{equation}
where $\widehat{\textbf{R}}$ and $\widehat{\textbf{x}}$ are the unit
vectors of the above-defined \emph{R} and \emph{x} directions.

We should mention, in passing, that the temperature structure in
x-direction could be modelled in various ways. We could choose the
Heaviside step function, for example, instead of $\sin^{2}(2\pi
x/\lambda)$  but refrained from assuming a jumpy passage to a 30$\%$
higher temperature from plumes to interplume lanes and vice versa.

\subsection{Number Densities in NPCH}
Electron number density distributions in the north and south polar
coronal holes are given by Fisher and Guhathakurta (1995). If the
values they report are typical for the plumes then the analytical
expression for electron number density distribution may be given as
below (Esser et al. 1999):
\begin{equation}
N_{\rm{e}}^{\rm{PL}}(R) =2.494\times 10^{6}R^{-3.76}+1.034\times 10^{7}R^{-9.64}+3.711\times 10^{8}R^{-16.86}\,\,\, \rm cm^{-3}
\end{equation}
where the superscript PL stands for plume. Marsch (1999), Endeve \&
Leer (2001) and Voitenko \& Goosens (2002) regard the coronal hole
plasma as \emph{quasineutral}, i.e., $N_{\rm{e}}\approx
N_{\rm{p}}\approx N$ . Raymond et al. (1997) obtained the abundances
of oxygen and other elements in coronal streamers. They found an
oxygen number density $N_{\rm{OVI}}=6.8\times10^{-5}N_{\rm{p}}$ for
the streamer center. Ciaravella et al. (1999) also observed a
coronal streamer and obtained
$N_{\rm{OVI}}=5.9\times10^{-4}N_{\rm{p}}$. In the model for the
kinetics of ions in the solar corona, Vocks (2002) chose the oxygen
density as $N_{\rm{OVI}}=10^{-3}N_{\rm{p}}$. Finally, in a recent
study, Cranmer et al. (2008) give the upper and lower values of O VI
number densities in polar coronal holes
$N_{\rm{OVI}}=2.4\times10^{-6}N_{\rm{p}}$ and
$N_{\rm{OVI}}=8\times10^{-7}N_{\rm{p}}$, respectively. Taking a
simple average they have a mean value of
$N_{\rm{OVI}}=1.52\times10^{-6}N_{\rm{p}}$. Since the observational
uncertainty on the O VI abundances in NPCH still persists, we'll use
the highest and the lowest values for O VI number density, i.e.,
$N_{\rm{OVI}}=10^{-3}N_{\rm{p}}$ and
$N_{\rm{OVI}}=1.52\times10^{-6}N_{\rm{p}}$ in our calculations in
Section 3 to see how sensitive the wave propagation characteristics
is to the O VI abundance. We should mention in passing that Cranmer
(2000), Cranmer et al. (1999) and Tu \& Marsch (1999) showed that
minor ion species can dissipate the ICWs faster than the more
abundant ones.

Cranmer et al. (1999) used the mean electron density in coronal
holes, in other words, they derived the analytical form of the
electron density by averaging over plumes and interplume lanes.
However, number densities of electrons in plumes appear to be 10\%
higher than that of interplume lanes (see, e.g. Kohl et al. 1997a).
Since we assume, in the light of observational data, that
$N_{e}\approx N_{p}$, then the number density of protons in
(\emph{R, x}) plane becomes (Devlen $\&$ Pek{\"u}nl{\"u}, 2010),
\begin{equation}
N_{\rm{p}}(R,x)=N_{\rm{p}}^{\rm{PL}}(R)[1-0.1\sin^{2}(2\pi x/\lambda)] \,\,\, \rm cm^{-3}
\end{equation}
Here the factor 0.1 comes from the difference of the number
densities between plumes and interplume lanes. If we should adopt
the \emph{quasineutral} assumption then the O VI number density may
be expressed as below. Subscript \emph{i} stands for O VI (Devlen
$\&$ Pek{\"u}nl{\"u}, 2010).
\begin{equation}
N_{\rm{i}}(R,x)=fN_{\rm{p}}^{\rm{PL}}(R)[1-0.1\sin^{2}(2\pi x/\lambda)] \,\,\, \rm cm^{-3}
\end{equation}
where \emph{f} will assume values $1.52\times10^{-6}$ as given by
Cranmer et al. (2008) or $10^{-3}$ as adopted by Vocks (2002) in his
model. For the purpose of detecting Alfven waves in PCHs through EUV
line width variations, Banerjee et al. (2009b) used EIS/Hinode
spectrometer. The authors studied the electron number density
variation with nonthermal velocity for PCH. An interested reader may
refer to their Figure 5 in the above-cited article.

\subsection{Magnetic Field in NPCH}
To the best knowledge of the authors of the present investigation,
there is no indication in  the literature of  this field, as to
whether the magnetic field shows any spatial variation in \emph{x}
direction, therefore heliocentric distance dependence of NPCH
magnetic field given by Hollweg (1999a) will be used in the next
section:
\begin{equation}
B=1.5(f _{\rm max}-1)R^{-3.5}+1.5 R^{-2} \,\,\rm Gauss
\end{equation}
where $f_{\rm max}=9$. This model is valid in the range 1.0 - 10.0 R
which covers the range we are interested in, i.e. 1.5 - 3.5 R. O VI
cyclotron frequency's ($\omega_{\rm c}=q_{\rm i}B/m_{\rm i}c$)
radial dependence will be calculated by using the Eq. (9).

\subsection{Ion - Cyclotron Waves (ICW) in PCH}
ICW may be generated by various mechanisms in various parts of the
PCH. We do not touch upon the wave generation mechanisms, but simply
take ICW for granted. Generation of resonant ICW may be possible by
stochastic magnetic foot point motions, magnetic reconnections and
MHD filamentation instabilities or from MHD turbulent cascade. This
latter mechanism is supposed to be the dominant one producing ICW
that heat the coronal hole plasma and accelerate the solar wind
particles (Cranmer 2000). Axford et al. (1999) suggest that these
waves may be generated by turbulent cascade from low to high
frequencies or directly. Gupta et al. (2010), using EIS/Hinode and
SUMER/SOHO data, detected propagating disturbances in coronal lines
in plume and interplume lanes. They concluded that the waves are
probably either Alfvenic or fast magnetoacoustic in the interplume
regions and slow magnetoacoustic in plumes. Tomczyk et al. (2007)
detected Alfven waves by using Coronal Multi-Channel Polarimeter
(CoMP) at the National Solar Observatory, New Mexico. Observations
showed the existence of upward propagating waves with phase velocity
$1-4\times10^{6}$m/s. They concluded that the waves are too weak to
heat the solar corona and added that the unresolved Alfven waves may
carry enough energy to heat the corona. Doorsselaere et al. (2007)
appraised the observation of Alfven waves by Tomczyk et al. (2007)
as an important new progress in coronal physics. The important point
in this context is as to whether the wave flux density of the ICW is
high enough to replace the energy lost by thermal conduction to the
transition region and the optically thin emission. A recent
evaluation of the wave energy flux is given by McIntosh et al.
(2011). By using the SDO/AIA image sequences and applying Monte
Carlo simulations, the authors measured the amplitudes, periods and
phase speeds of transverse waves. They concluded that the estimated
energy flux in the quiet corona and coronal holes could supply the
energy needed for fast solar wind in two stages: at the first stage,
injected plasma from lower part of the atmosphere heats the coronal
base and at the second stage Alfvenic waves dissipate and accelerate
the solar wind in PCH.

Energy flux density of the ICW is given as (e.g., Banerjee et al. 1998),
\begin{equation}
F_{\rm \omega}=\sqrt{\rho/4\pi}\langle\delta v^{2}\rangle B \,\,\,\,\rm erg cm^{-2} s^{-1}
\end{equation}
The wave amplitude at heights $120''$ off the solar limb is about,
$\langle\delta v^{2}\rangle=2\times(43.9\,\rm km s^{-1})^{2}$.
Adopting the values for B = 5 G and $N_{\rm e} = 4.8\times10^{13}\,
\rm m^{-3}$ at r = 1.25 R, Banerjee et al. (1998) found the wave
flux density as $F_{\rm \omega}= 4.9\times10^{5}\, \rm erg
cm^{-2}s^{-1} $ which is high enough for the ion-cyclotron resonance
(ICR) process to be a good candidate for heating the coronal hole.

We may assume that the most favorable process for heating the NPCH
is the ICR process. We may justify this assumption on the ground of
two observational facts: a) Doppler dimming analysis (Kohl et al.
1997a) showed that the perpendicular (to the magnetic field)
temperature of the O VI ions is about two order of magnitude higher
than the parallel one, i.e., $T_{\bot}\sim 10^{2}T_{\|}$  and b)
outflow velocities in the solar wind of O VI ions in the interplume
lanes are higher than that of the lanes. The seat of the fast solar
wind was identified as the interplume lanes (Wilhelm et al. 1998;
Hollweg 1999a, 1999b, 1999c). This shows that the ions having
greater perpendicular velocities will experience a higher mirror
force, $F_{\rm mf}(R)=-\mu\nabla B= -[(1/2) m_{\rm i}
v^{2}_{\bot}/B(R)]\nabla B $. These two effects could be brought
about only by ICR process (Hollweg 1999a, 1999b, 1999c).

Therefore the ICR process is regarded as the most efficient heating
mechanism for PCH. Observations also show that whatever the heating
mechanism, O VI ions are preferentially selected (Kohl et al.
1997b). Alfven waves undergoing the ICR process and dissipating
their energies was examined by Cranmer (2000). One has to refer to
this paper for earlier work in this field. In Section 3 we present
our model based on the observational results presented in this
section.

\section{Solution of the Vlasov Equation in PIPL structure of PCH}

We consider the wave equation derived for a cold plasma. In order to
justify the \emph{cold plasma} approximation, we should point out
that two conditions are to be fulfilled in PCH: a) perpendicular
wavelength should be greater than O VI ion Larmor radius
($k_{\bot}v_{\bot}/\omega_{c}\ll 1$) and b) parallel phase velocity
of ICW should be larger than the O VI ion thermal velocity
($v_{\parallel}\ll \omega/k_{\parallel}$) (Schmidt, 1979). We assume
a quasi-parallel propagation, in the sense that
$k_{\bot}/k_{\parallel}\ll1$, in PCH, therefore the former condition
is readily fulfilled. The latter condition will be shown satisfied
\emph{a posteriori}. We assume that the perturbed quantities
variation in space and time is like a plane wave, i.e.,
$\exp[i(\emph{\textbf{k}}\cdot\emph{\textbf{r}}-\omega t)]$, where
\textbf{\emph{k}} is the wave vector, \textbf{\emph{r}} is the
distance from the source, $\omega$ is the wave frequency and
\emph{t} is the time. In our investigation, we'll not take the
$\cos\theta$ factor in $\textbf{k}\cdot \textbf{r}=\mid
\textbf{k}\mid\mid\textbf{r}\mid\cos\theta$ , where $\theta$ is the
angle between the wave vector and the magnetic field. It is implicit
in the above given condition, i.e., $k_{\bot}/k_{\parallel}\ll1$,
that we may assume a quasi - longitudinal propagation but not
forgetting the fact that we need $k_{\bot}$ so that the refraction
of the waves can cause communication between the adjecent flux tubes
in plumes and interplume lanes. Murawski et al. (2001) also modeled
the coronal hole as a cold plasma slab but with a uniform magnetic
field. Written in terms of Fourier components, the wave equation is
as below (e.g., Stix 1962, 1992),
\begin{equation}
\textbf{k}\times \left(\textbf{k} \times \textbf{E}\right)+\frac{\omega^{2}}{c^{2}} \kappa \cdot \textbf{E}=0
\end{equation}
where $\kappa$ is the dielectric tensor and will be derived from the
Vlasov equation. O VI ions will be considered under the two forces,
a) Lorentz force,
$q[\textbf{E}+\frac{1}{c}(\textbf{v}\times\textbf{B}_{0})]$ where
$\textbf{E}$ is the electric field of the ICW waves and
$\textbf{B}_{0}$ is the external magnetic field of the PCH and b)
pressure gradient force, $-\nabla p$.

Before proceeding with the linearization of the Vlasov equation we
should discuss the justifiability of the pressure gradient term in
the Vlasov equation.

The Vlasov equation usually does not contain the pressure gradient
term. More recently, Tronci (2010) mentions so-called
\textquotedblleft hybrid kinetic-fluid models\textquotedblright and
argues that, \textquotedblleft These models are usually realized by
following a hybrid philosophy that couples ordinary fluid models to
appropriate kinetic equations governing the phase-space distribution
of the energetic particle species\textquotedblright (see Eq. 70 in
Tronci's paper).

Cheng (1991) argues that plasmas in a large magnetic fusion devices
as well as in space can be considered as having two components: the
background one with a low energy which satisfies the fluid
description (MHD) and the energetic one with a low density which
should be treated by the kinetic approach. Cheng (1999) further
proposed the extension of the kinetic-MHD model by properly
including important kinetic effects of all the plasma species.

Another example reveals itself in the astrophysical literature
dealing with the plasma instabilities in clusters of galaxies
(Schekochihin et al., 2005).  The authors of this study include the
pressure gradient term in the Vlasov equation (see their Eq. 5) and
cite Kulsrud (1983). This paper is different from Tronci's (2010)
one in the sense that there are no velocity moment equations (i.e.,
MHD equations) coupling to the Vlasov equation.

Cranmer (2002) identifies the thermal pressure gradient in the hot
corona as the dominant outward force on particles in both fluid and
kinetic models and notes that \emph{\textquotedblleft this force is
a completely collisionless phenomenon\textquotedblright}. This
unconventional description of pressure gradient force combined with
the observational data clearly show that both the density and the
temperature of O VI ions are the functions of radial distance from
the Sun and along the x - direction. Thermal pressure gradients are
very steep both in the radial and the x - direction in NPCH; so are
the density gradients, although little milder.

The reasons why we did not consider the pressure force from the
background plasma are, i) collisonless nature of the NPCH is well
established and ii) NPCH fulfills the non-neutral plasma and One
Component Plasma (OCP) conditions, i.e., $r_{c}\ll \lambda_{D}$ and
$L_{T}\geq \lambda_{D}$; where $r_{c}$ is the Larmor radius of O VI
ions, $\lambda_{D}$ is the Debye length and $L_{T}$ is the
temperature length scale across the magnetic field (Dubin \&
O'Neill, 1997).

The crucial point is if the PCH could be regarded as a non-neutral
plasma. Dubin \& O'Neill (1999) points to the similarity between
non-neutral plasma and OCP. Particle beams are used as a new
application in non-neutral plasma experiments (Marler \& Stoneking,
2007). O VI ions are preferentially heated species of NPCH plasma.
Their effective temperature reaches to $10^8$ K. This is at least
two orders of magnitude higher than those of any other plasma
species in the range 1.6 - 3.5 R. O VI ions may be regarded as
\textquotedblleft ion beams\textquotedblright in the quasi-neutral
and immobile (with respect to O VI ions) NPCH background plasma as a
non-neutral OCP.

We do not want to mislead the reader with the concept of
\textquotedblleft mobility\textquotedblright. It is a well
established fact that NPCH is collisionless. Mobility is meaningful
in a collisional plasma. In a neutral or quasi-neutral plasma
electrons are more mobile, therefore in such media electron Debye
length is used. On the other hand, in a collisionless plasma,
\textquotedblleft collision frequency\textquotedblright,
\textquotedblleft mean free path\textquotedblright and
\textquotedblleft collision time scale\textquotedblright appearing
in the formula of the mobility lose their validities.

If we can consider the NPCH as a non-neutral OCP then only the ion
Debye length should be used in the above two conditions,i.e.,
$r_{c}\ll \lambda_{D}$ and $L_{T}\geq \lambda_{D}$. Because electron
Debye length is formulated under several assumptions: a) plasma is
in a thermodynamic state; b) in case electron Debye length is
considered then shielding is due to only one sign, i.e., electrons.
This is a cold ion approximation ($T_{\rm i}$ = 0) wherein $T =
T_{\rm e}$ is valid (Somov, 2006). Neither of the above two
assumptions are justifiable in the NPCH. SoHO observations showed
that more than 2000 plasma species co-exist with different
temperatures in NPCH, besides their parallel and perpendicular (to
the external magnetic field) velocities are quite different. Again,
SoHO observations revealed that  $T_{\rm e} \sim 10^5-10^6$ K but
$T_{\rm i}^{OVI}\sim 5\times 10^{6}-10^{8}$K. In NPCH, electrons are
too slow to shield O VI ions. With these observational data in hand,
we may claim that short range collisions could occur only between O
VI ions themselves, if at all.

Through the radial distance 1.6-3.0 R, $r_{c}/\lambda_{D}$ changes
from 0.061 to 0.095; $L_{T}/\lambda_{D}$ changes from $7\times10^4$
to $1.81\times10^4$, respectively. Therefore the quantitative
criteria,i.e., $r_{c}\ll \lambda_{D}$ and $L_{T}\geq \lambda_{D}$,
are fulfilled in the NPCH.

The  $\nabla n$ and  $\nabla T$ terms we consider are exclusively
for the O VI ions, i.e., O VI ion number density as well as the
perturbed part of the effective temperature variations in the radial
and x - directions. Bearing all these in mind, we follow the
Schekochihin et al. (2005) model and insert the pressure gradient
term into the Vlasov equation. In this case the quasi-linearized
Vlasov equation becomes,
\begin{equation}
\begin{array}{l}\frac{{\it d}\, {\it f}_{1}}{{\it d}\, {\it t}} =\frac{\partial \, {\it f}_{1}}{\partial \, {\it t}} +\textbf{v}\cdot \frac{\partial \, {\it f}_{1} }{\partial \, \textbf{R}} +\frac{{\it q}_{\rm i}}{{\it m}_{\rm i}} \left(\textbf{E}_{1} +\frac {1}{c}(\textbf{v}\times \textbf{B}_{0}) \right)\cdot \frac{\partial \, {\it f}_{1}}{\partial \, \textbf{v}} = \bigg[ -\frac{{\it q}_{\rm i}}{{\it m}_{\rm i} } \left(\textbf{E}_{1} +\frac {1}{c}(\textbf{v}\times \textbf{B}_{1}) \right) \\
\\
+\frac{{\it k}_{ \rm B}}{n_{0}m_{\rm i}} \left(T_{\rm eff}^{\rm \xi}\frac{\partial n_{\rm 0}}{\partial \textbf{R}}+T_{\rm eff}^{\rm \xi}\frac{\partial n_{0}}{\partial \textbf{x}}+
n_{0}\frac{\partial T_{\rm eff}^{\rm \xi}}{\partial \textbf{R}}+n_{0}\frac{\partial T_{\rm eff}^{\rm \xi}}{\partial \textbf{x}}\right) \bigg]\cdot \frac{\partial \, {\it f}_{0} }{\partial \textbf{v}} \end{array}
\end{equation}
where $f_{0}$ and $f_{1}$ are the unperturbed and perturbed parts of
the velocity distribution function, and $\textbf{E}_{1}$ and
$\textbf{B}_{1}$ are the wave electric and magnetic fields,
respectively. The time derivative is taken along the unperturbed
trajectories in phase - space of O VI ions. Hereafter, for the sake
of brevity, we'll designate the linearized form of the pressure
force as,
\begin{equation}
 \nabla p_{1}=\frac{{\it k}_{ \rm B}}{n_{0}m_{\rm i}} \left(T_{\rm eff}^{\xi}\frac{\partial n_{0}}{\partial \textbf{R}}+T_{\rm eff}^{\xi}\frac{\partial n_{0}}{\partial \textbf{x}}+
n_{0}\frac{\partial T_{\rm eff}^{\xi}}{\partial \textbf{R}}+n_{0}\frac{\partial T_{\rm eff}^{\xi}}{\partial \textbf{x}}\right)
\end{equation}

Left circularly polarized ICW which interacts with the perturbed
velocity distribution function is given by Schmidt (1979). Since we
extend his study by taking the $\nabla p$ term into account, the
perturbed velocity function is modified as below,
\begin{equation}
\begin{array}{l}{f_{\rm L}=}
 {-\frac{q_{\rm i} }{m_{\rm i} } \left[\left(1-\frac{v_{\parallel} k}{\omega } \right)\frac{\partial \, f_{0} }{\partial \, v_{\bot } } +\frac{k\, v_{\bot } }{\omega } \frac{\partial \, f_{0} }{\partial \, v_{\parallel} } \right]E_{\rm x} \exp \, \left(i\theta _{0} \right)\frac{1-\exp \phi}{i(k\, v_{\parallel} -\omega + \omega _{c} )} +\nabla p_{1}}
 \end{array}
\end{equation}
where  $\theta_{0}$ is the initial phase of the wave and
$\phi=\left[i\left(k\, v_{\parallel} -\omega+\omega _{c}
\right)\left(t-t_{0} \right)\right]$. $f_{1}$  produces a current
the \emph{x} - component of which is (\emph{ibid}),
\begin{equation}
\begin{array}{l} {J_{\rm x} =- \frac{q_{\rm i}^{2} \, \pi }{m_{\rm i} \omega } \, \, E_{\rm x} \int \frac{\left(\omega -k\, v_{\parallel} \right)\, \left(\partial \, f_{0} /\partial \, v_{\bot } \right)+k\, v_{\bot } \left(\partial \, f_{0} /\partial \, v_{\parallel} \right)}{k\, v_{\parallel} -\omega + \omega _{c} } (1-\exp \phi ) v_{\bot }^{2} dv_{\bot } \, dv_{\parallel} } \\
\\
 { +i\pi {\it q}_{{\rm i}} \int \nabla p_{1} (t-t_{0} ) v_{\bot }^{2} \, dv_{\bot } \, dv_{\parallel} } \end{array}
\end{equation}
$J_{\rm y}$ component of the current is found in a similar manner.
The result is $J_{\rm x}/E_{\rm x}=J_{\rm y}/E_{\rm
y}=\textbf{J}/\textbf{E}$ . The plasma dielectric tensor is obtained
from the relation given below (e.g. Stix 1962)
\begin{equation}
\textbf{J}+ \frac{i\omega}{4\pi}\textbf{E}= \frac{i\omega}{4\pi}\kappa\cdot \textbf{E}
\end{equation}
Using Eq. (16) we obtain the dielectric tensor for the left
circularly polarized ICW as below:
\begin{equation}
\begin{array}{l} { \kappa_{L} =1+4\pi \frac{J_{\rm x} /E_{\rm x} }{i\, \omega \ } =} {1-
 \frac{4\pi^{2} \,q_{\rm i}^{2}  }{im_{\rm i}\ \omega ^{2} } \int _{-\infty }^{+\infty }d\, v_{\parallel} \int _{0}^{\infty }\frac{\left(\omega -k\, v_{\parallel} \right)\, \left(\partial \, f_{0} /\partial \, v_{\bot } \right)+k\, v_{\bot } \left(\partial \, f_{0} /\partial \, v_{\parallel} \right)}{k\, v_{\parallel} -\omega + \omega _{c} } (1-\exp \phi ) v_{\bot }^{2} \, d\, v_{\bot }   } \\
\\
+ {\frac{4\pi^{2} q_{i} }{ \omega E_{\rm x} } (t-t_{0} )\int _{0}^{\infty } \nabla p_{1} v_{\bot }^{2} \, dv_{\bot } \, dv_{\parallel} } \end{array}
\end{equation}
The dispersion relation for left circularly polarized ICW is
obtained by $\kappa_{L}=n^{2}=c^{2}k^{2}/\omega^{2}$, where n is the
refractive index of the medium for ICW.

Bearing in mind the collisionless nature of the PCH and the
observational fact that $T_{\bot}\sim 10^{2}T_{\parallel}$  we
assumed that the velocity distribution function is bi-Maxwellian
(see also Cranmer et al. 2008),
\begin{equation}
f_{0} =n_{\rm i} \, \alpha _{\bot }^{2} \alpha _{\parallel} \, \pi ^{-3/2} \exp \left[-\left(\alpha _{\bot }^{2} v_{\bot }^{2} +\alpha _{\parallel}^{2} v_{\parallel}^{2} \right)\right]
\end{equation}
where $\alpha_{\bot}=\left(2k_{\rm B}T_{\bot}/m_{i}\right)^{-1/2}$
and $\alpha_{\parallel}=\left(2k_{\rm
B}T_{\parallel}/m_{i}\right)^{-1/2}$ are the inverse of the most
probable speeds in the perpendicular and parallel direction to the
external magnetic field, respectively. $\partial f_{0}/\partial
v_{\bot}$ and $\partial f_{0}/\partial v_{\parallel}$ appearing in
Eq. (17) will be derived from Eq. (18). We should remind the reader
that $ [\textbf{E}+\frac{1}{c}\textbf{v}\times \textbf{B}_{0}-\nabla
p] \cdot\nabla_{v}f$  term in the Vlasov equation is a non-linear
term, therefore harbours non-linear effects like the ICR process.
This property of the velocity distribution function makes the
approximation quasi-linear. We may justify this assumption on the
ground that, quasi-linear theory provides an accurate description
for small amplitude waves. In this small amplitude limit, exchange
of energy between waves and particles are well quantified (e.g.
Goldston \& Rutherford 1995; Gurnett \& Bhattacharjee 2005). Cranmer
(2000) also considered small amplitude waves in a collisionless,
homogeneous plasma wherein he points to the fact that in this ideal
case dissipation of ICW occurs via ICR. In this investigation, we
depart from the homogenous case and consider all the gradients of
plasma parameters which are observationally revealed. When we
substitute the $\partial f_{0}/\partial v_{\bot}$ and $\partial
f_{0}/\partial v_{\parallel}$ into Eq. (17) we get the contribution
of the principle integral to the dispersion relation of ICW as
below,

\begin{equation}
\begin{array}{l} k^{2} \, \left(c^{2} -\frac{\omega^{2} }{k^{2} } \right)-\frac{i\omega_{p}^{2} \, \omega}{\left(\omega -\omega_{c} \right)} -\frac{i\omega _{p}^{2} \, k}{\sqrt{\pi } \, \alpha_{\parallel} \, \left(\omega -\omega_{c} \right)} \left(\frac{T_{\bot } }{T_{\parallel} } +\frac{\omega }{(\omega -\omega _{c})}-1\right) \\
\\
 -\frac{i\omega _{p}^{2} \, k^{2} }{2\alpha_{\parallel}^{2} \, \left(\omega -\omega_{c} \right)^{2} } \left(\frac{\omega }{(\omega -\omega_{c} )} +\frac{T_{\bot } }{T_{\parallel} } -1\right)+\frac{\pi^{2}\omega {\it q}_{i}\, L }{{\it E}_{\rm x} v_{A}}\nabla p_{1}^{\rm r}=0
 \end{array}
\end{equation}

where $\omega_{p}=(4\pi n_{OVI}q_{i}^{2}/m_{i})^{1/2}$ is the plasma
frequency for the O VI ions and $\nabla p_{1}^{r}$ is the term
designating rather lengthy pressure gradient in its reduced form:
\begin{equation}
\nabla p_{1}^{\rm r}=\frac{{\it k}_{ \rm B}}{m_{\rm i}}\left[{ \left(T_{\rm eff}^{\xi}\frac{\partial n_{0}}{\partial R}+n_{0}\frac{\partial T_{\rm eff}^{\xi}}{\partial R}\right) \left(\frac{T_{\bot}}{T_{\parallel}}\right)^{1/2}}+2\left(T_{\rm eff}^{\xi}\frac{\partial n_{0}}{\partial x}+n_{0}\frac{\partial T_{\rm eff}^{\xi}}{\partial x}\right)\right]
\end{equation}
We put  $L/v_{\rm A}$ instead of the time interval term $t-t_{\rm
0}$ in Eq. 17, where \emph{L} is the displacement of a wave (3.5-1.5
R) and $v_{\rm A}$ is the Alfven speed, i.e., $v_{\rm A}=B(R)/(4 \pi
n_{\rm i}m_{\rm i})^{1/2}$. Hereafter, for the sake of brevity we
replace $[\omega /(\omega -\omega _{c} )] +(T_{\bot }
/T_{\parallel}) -1$  by $\Pi$ which stands for the principle
contribution.

The residual contribution is:
\begin{equation}
\Re=\frac{2\sqrt{\pi } \omega _{p}^{2} }{k} \, \alpha _{\parallel} \left[\omega _{c} +(\omega -\omega _{c} )\frac{T_{\bot } }{T_{\parallel} } \right]\exp \left[-\alpha _{\parallel} ^{2} \left(\frac{\omega -\omega _{c} }{k} \right)^{2} \right]
\end{equation} \\
Combination of the Eqs (19) and (21) is the dispersion relation for the ICW in PCH,
\begin{equation}
\begin{array}{l} {k^{2} \, \left(c^{2} -\frac{\omega ^{2} }{k^{2} } \right)-\frac{i\omega _{p}^{2} \, }{\left(\omega -\omega _{c} \right)} \left[\omega +\frac{k}{\sqrt{\pi } \, \alpha _{\parallel} \, } \Pi+\frac{\, k^{2} }{2\alpha _{\parallel}^{2} \, \left(\omega -\omega _{c} \right)} \Pi\right]}\\
\\
{+\frac{\pi^{2}\omega {\it q}_{{\it i}}^{} \, L } {{\it E}_{{\rm x}} v_{A}} \nabla p_{1}^{\rm r}

-\frac{2\sqrt{\pi } \omega _{p}^{2} }{k} \, \alpha _{\parallel} \Phi\exp \left[-\alpha _{\parallel} ^{2} \left(\frac{\omega -\omega _{c} }{k} \right)^{2} \right]=0} \end{array}
\end{equation}\\
where
$\Phi=\left[\omega_{c}+(\omega-\omega_{c})T_{\bot}/T_{\parallel}\right]$.
If we expand the exponential factor appearing in the last term of
the Eq. (22) into Taylor series, we obtain the \emph{dispersion
relation I} (DR I) as below,
\begin{equation}
\begin{array}{l} {k^{5} \left[{\it c}^{2} -\frac{{\it i}\omega _{p}^{2} \, }{2\alpha _{\parallel}^{2} \left(\omega -\omega _{c} \right)^{2} }\Pi \right]-k^{4} \frac{{\it i}\omega _{p}^{2} \, }{\sqrt{\pi } \, \alpha _{\parallel} \left(\omega -\omega _{c} \right)}\Pi
 -k^{3} \left[\omega ^{2} +\frac{{\it i}\omega _{p}^{2} \, \omega }{\left(\omega -\omega _{c} \right)} -\frac{\pi^{2}\omega {\it q}_{{\it i}} \, L } {{\it E}_{{\rm x}} v_{A} }\nabla p_{1}^{\rm r}\right]} \\
\\
{-k^{2} 2\sqrt{\pi } \omega_{p}^{2} \, \alpha_{\parallel} {\Phi+2}\sqrt{\pi } \omega_{p}^{2} \, \alpha_{\parallel}^{3} \left(\omega -\omega_{c} \right)^{2} {\Phi}=0} \end{array}
\end{equation}

\begin{figure}
\centering
\includegraphics{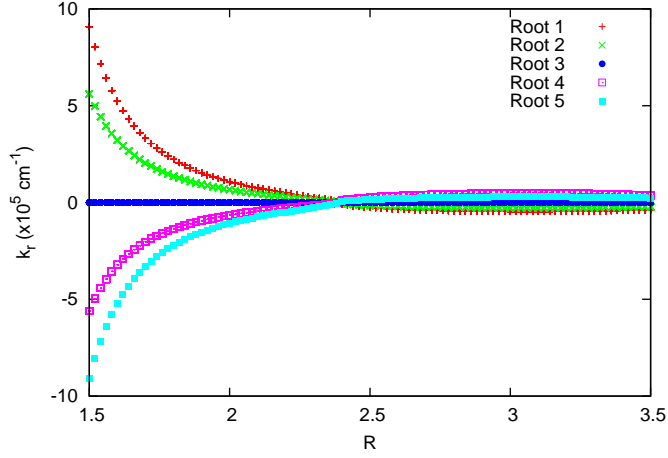}
\caption{All five roots of the fifth order dispersion relation given by Eq. 23. Roots 1, 2, 4 and 5 represent the forward propagating modes, and root 3 is the backward propagating mode.}
\end{figure}

\begin{figure}
   \centering
   \includegraphics[angle=270,scale=0.33]{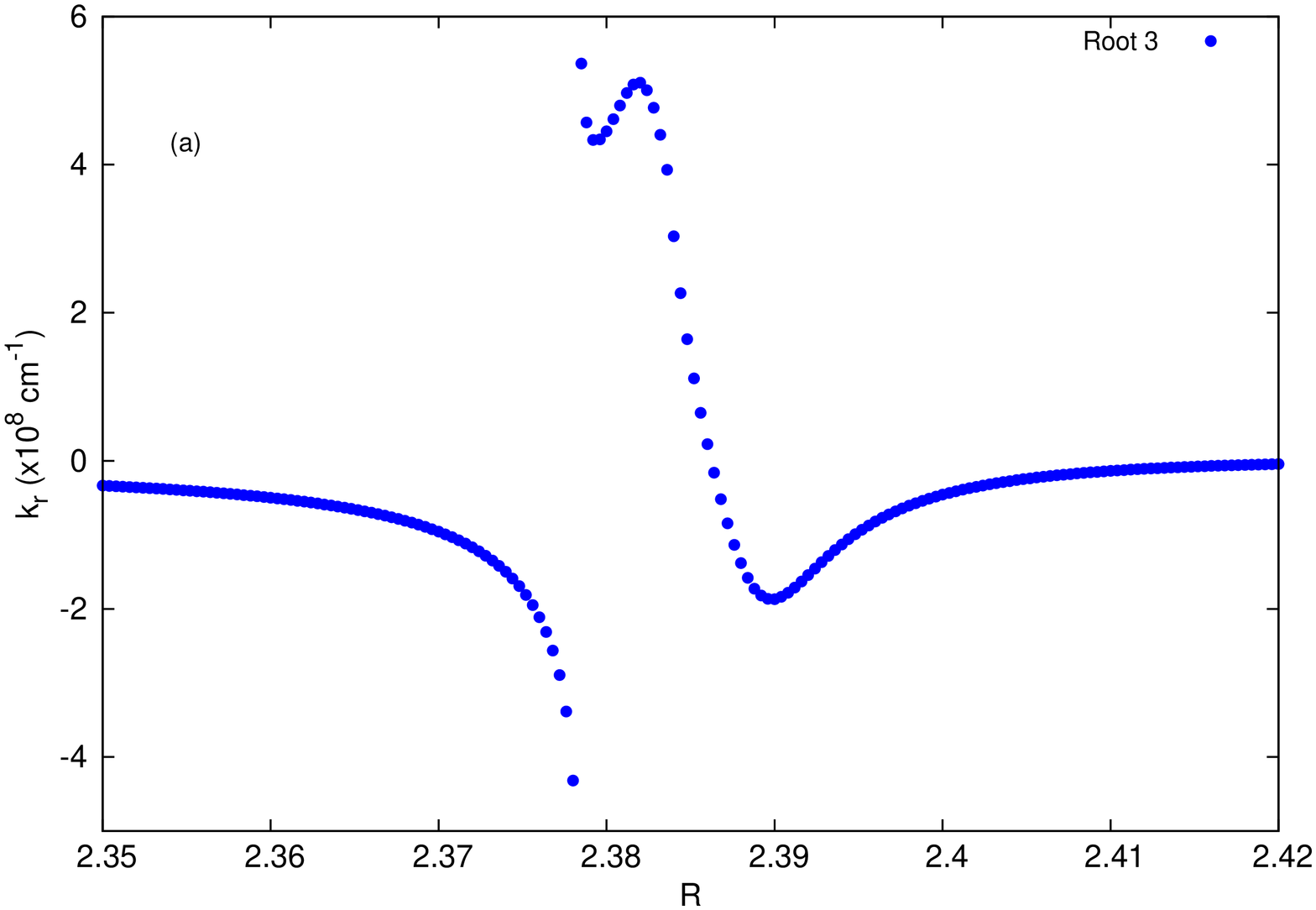}
   \includegraphics[angle=270,scale=0.33]{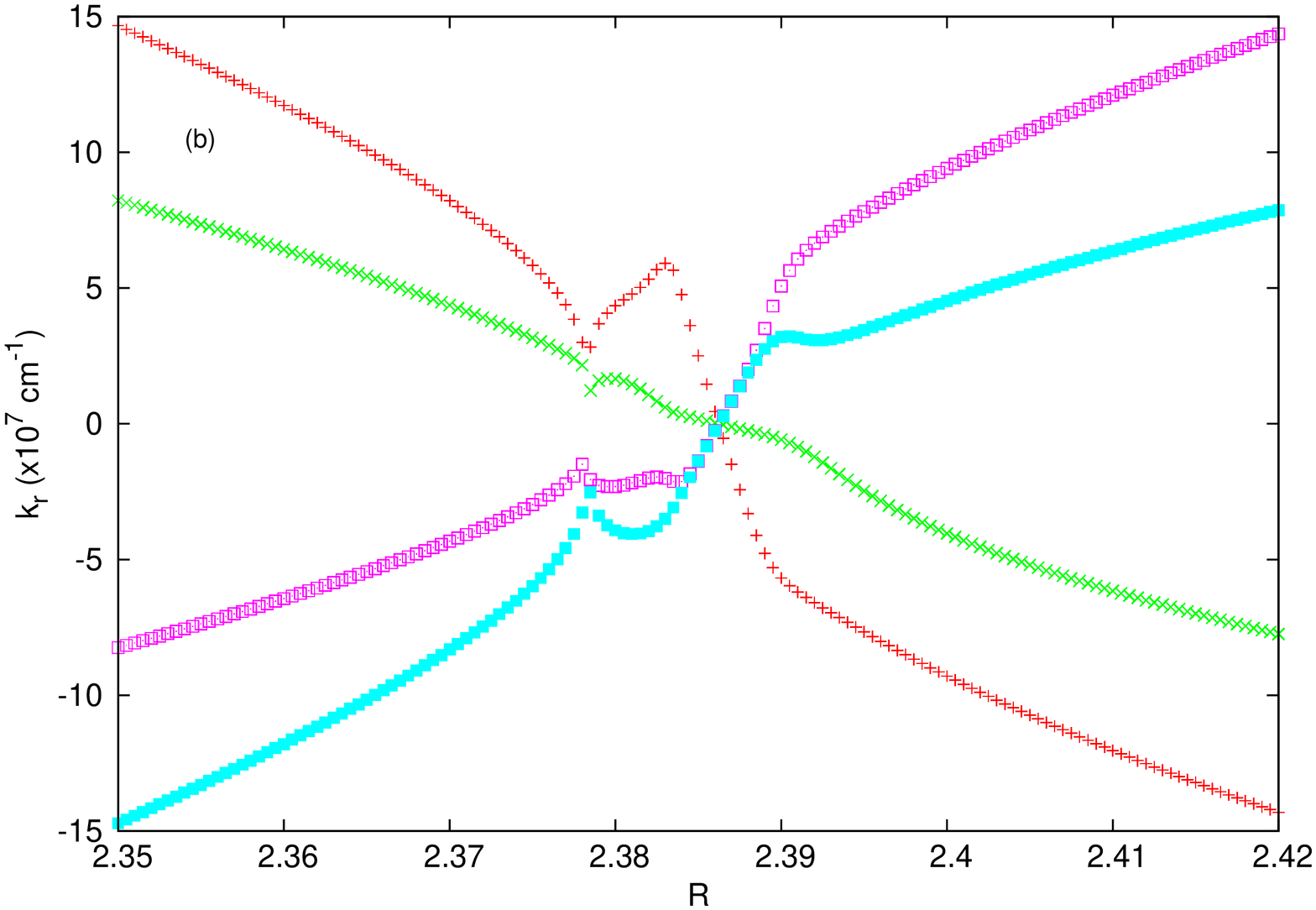}
\caption{(a) Root 3 only. This root represents the mode resonating with OVI ions at about R = 2.38; (b) Roots 1,2,4 and 5 in the reflection region. In the close vicinity of R = 2.39 fast and slow modes are reflected, i.e. wavenumbers become zero. The graphs are for ICW with a frequency 2500 $\rm rads^{-1}$.}
    \end{figure}

The graphical solution of the Eq. (23) (DR I) is given in Figure 2.
Whitelam et al. (2002) investigated the theoretical models that may
explain the formation of small scale jet-like structures known as
spicules and macrospicules in the solar transition region. In their
ponderomotive force acceleration model, they found that in order to
achieve accelerations of the order of 1 $\rm kms^{-2}$, an Alfven
wave with a field strength of order E $\approx$ 8 $\rm Vm^{-1}$ is
required (see their Eq. 16). We use this value for the $E_{\rm x}$
term in our equations. Terradas et al. (2010) investigated the
resonant absorption of the fundamental kink mode in a coronal loop.
They showed that the system becomes unstable when the frequencies of
the forward and the backward propagating waves merge. We obtained a
similar result, although our DR is solved for real frequency and
complex \emph{k}, the wave number. Waves in a lossy medium are
identified with respect to the analytical form of complex valued
wave numbers, i.e., $k=k_{r}+ik_{i}$. We assumed that the field
function of the wave is of the form, $exp[i(\textbf{k}\cdot
\textbf{r}-\omega t)]=exp[i(\textbf{k}_{r} \cdot \textbf{r}-\omega
t)]exp(-\textbf{k}_{i}\cdot \textbf{r})$. If the solution of the
dispersion relation yields $ k_{i} > 0 $ then the wave damps in the
$\textbf{r}$ -direction. Damping occurs if the energy flux of the
wave is also in the $\textbf{r}$ -direction. Shevchenko (2007)
introduces a criterion that helps to identify the waves, that is,
$Im k^{2}=2k_{r} k_{i} \gtrless 0$ where the upper (lower) sign is
for the backward (forward) wave. Our dispersion relation has five
roots, 4 of which is forward and the one of which is backward.

Figure 2 shows that forward and backward propagating fast and slow
modes with a frequency of $2500\,\rm rads^{-1}$ merge at about 2.38
R. We'll show in the Figure 3 that this location is the site where
ion cyclotron resonance and cut-offs take place in a small range of
distance. In Fig. 3a only the third root of DR I is shown. This root
reveals a resonance at about R = 2.38. The other four roots (Root
1,2,4 and 5) are given in Fig 3b. A reflection of the slow and fast
modes occurs at R=2.39. We should emphasize that the reflection site
and the resonance site are not coincident.

We should also remind the reader that the Taylor series
approximation is valid only in the close vicinity of the resonance
region, i.e. R=2.38 for the frequency $\omega = 2500 \rm rads^{-1}$.
Therefore our approximation for DR I would be invalid in the regions
away from the resonance.

\begin{figure*}
  \begin{center}
    \begin{tabular}{cc}
      {\includegraphics[angle=270,scale=0.28]{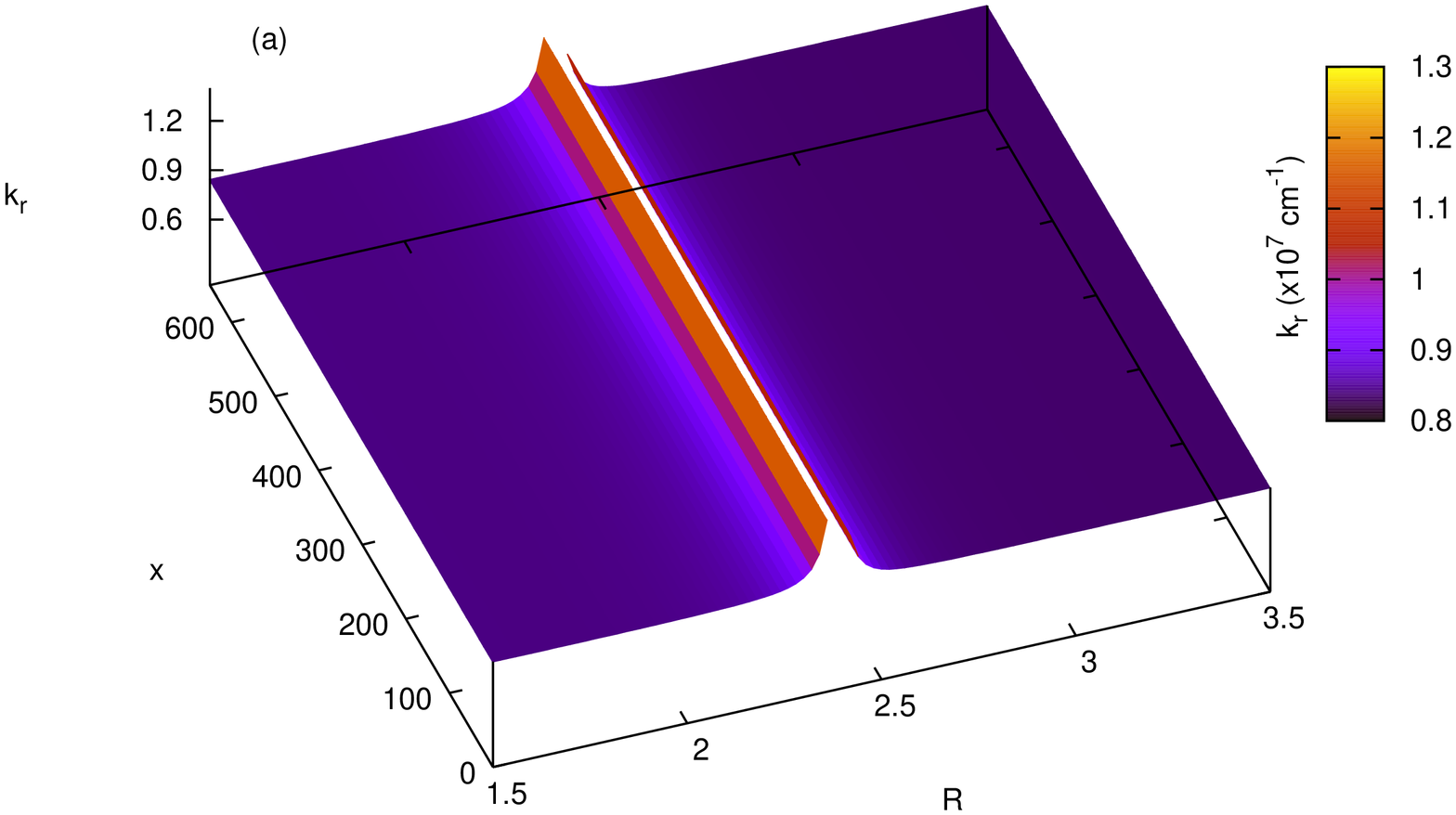}} &
      {\includegraphics[angle=270,scale=0.3]{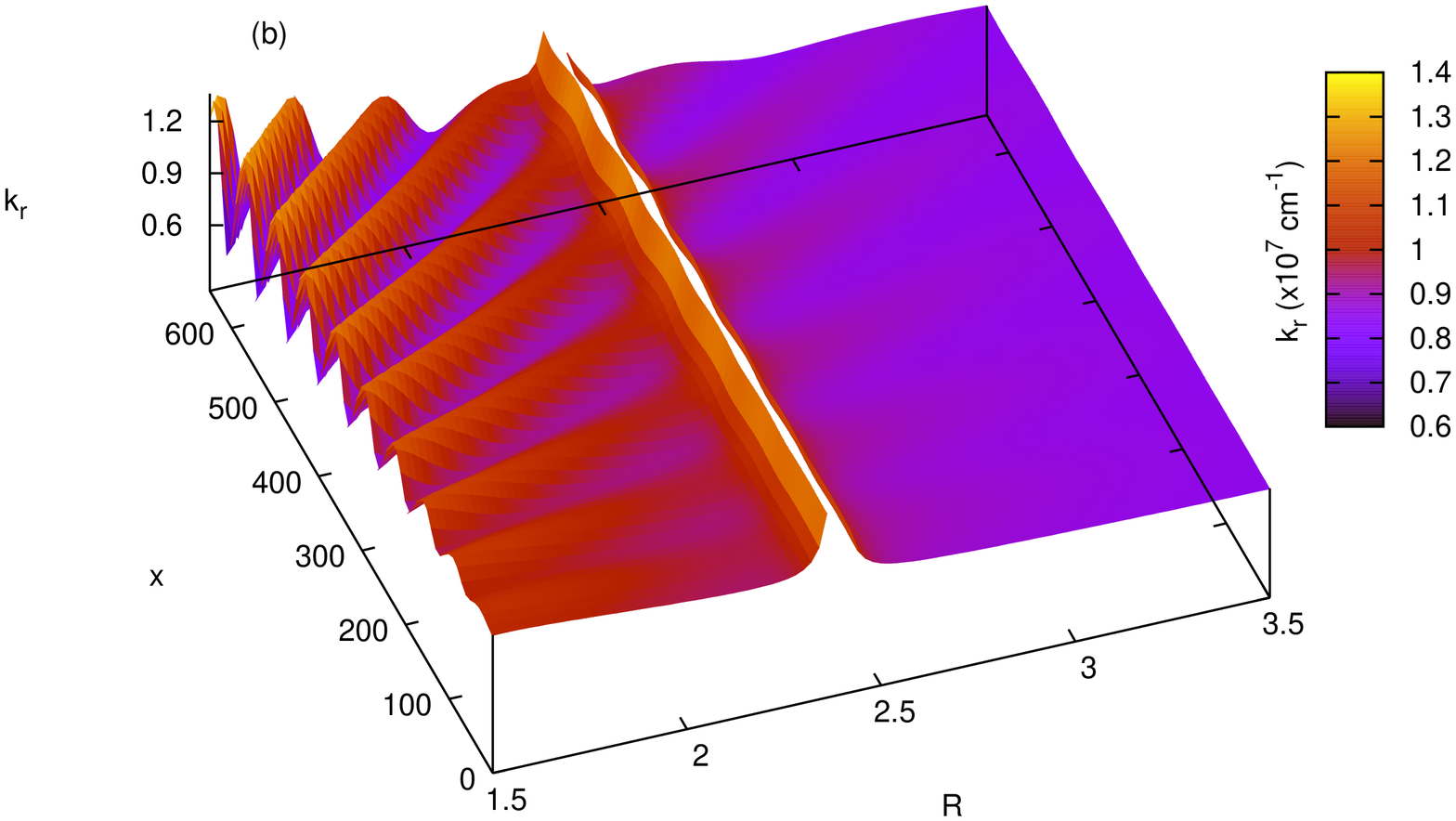}} \\
      {\includegraphics[angle=270,scale=0.28]{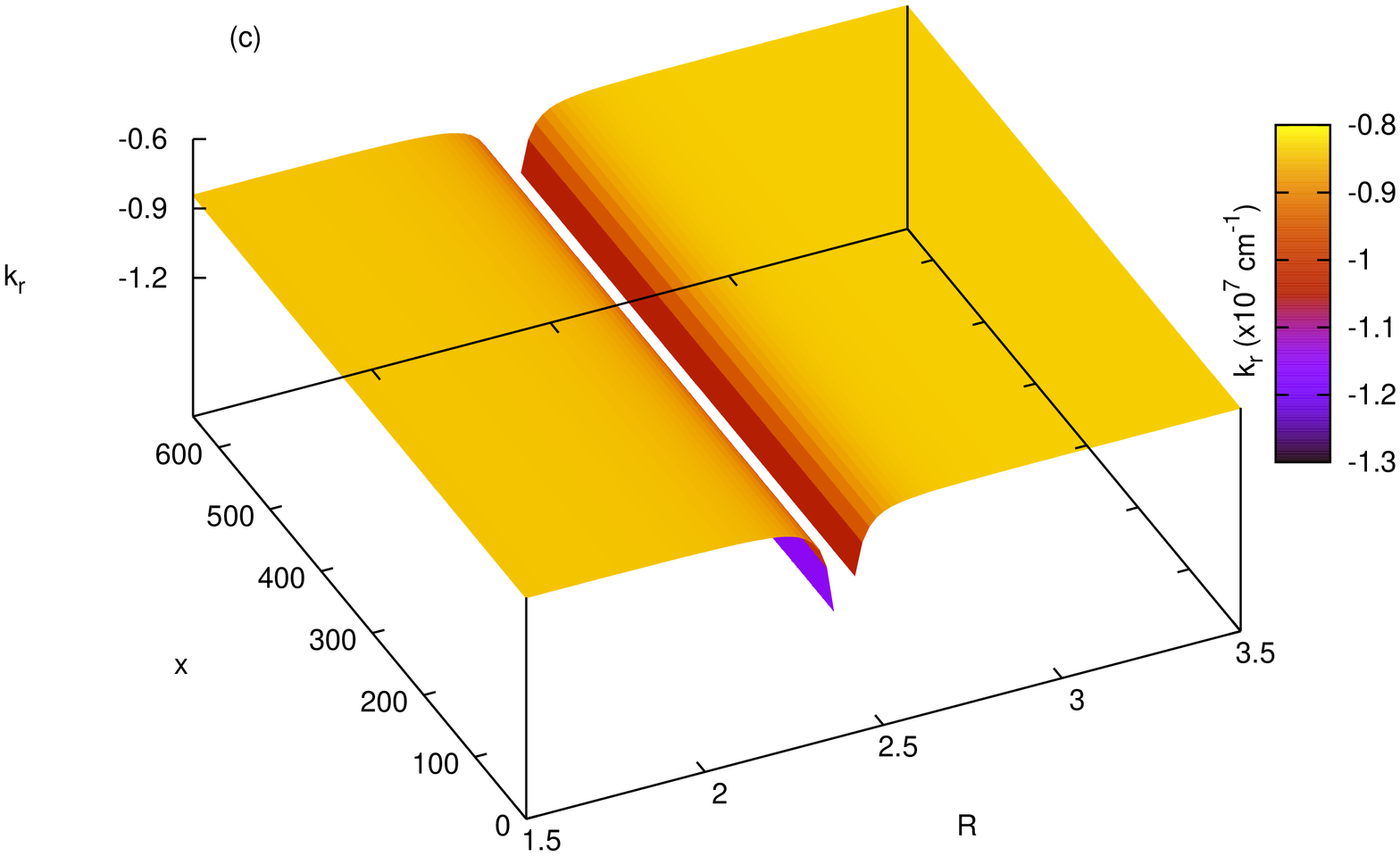}} &
      {\includegraphics[angle=270,scale=0.30]{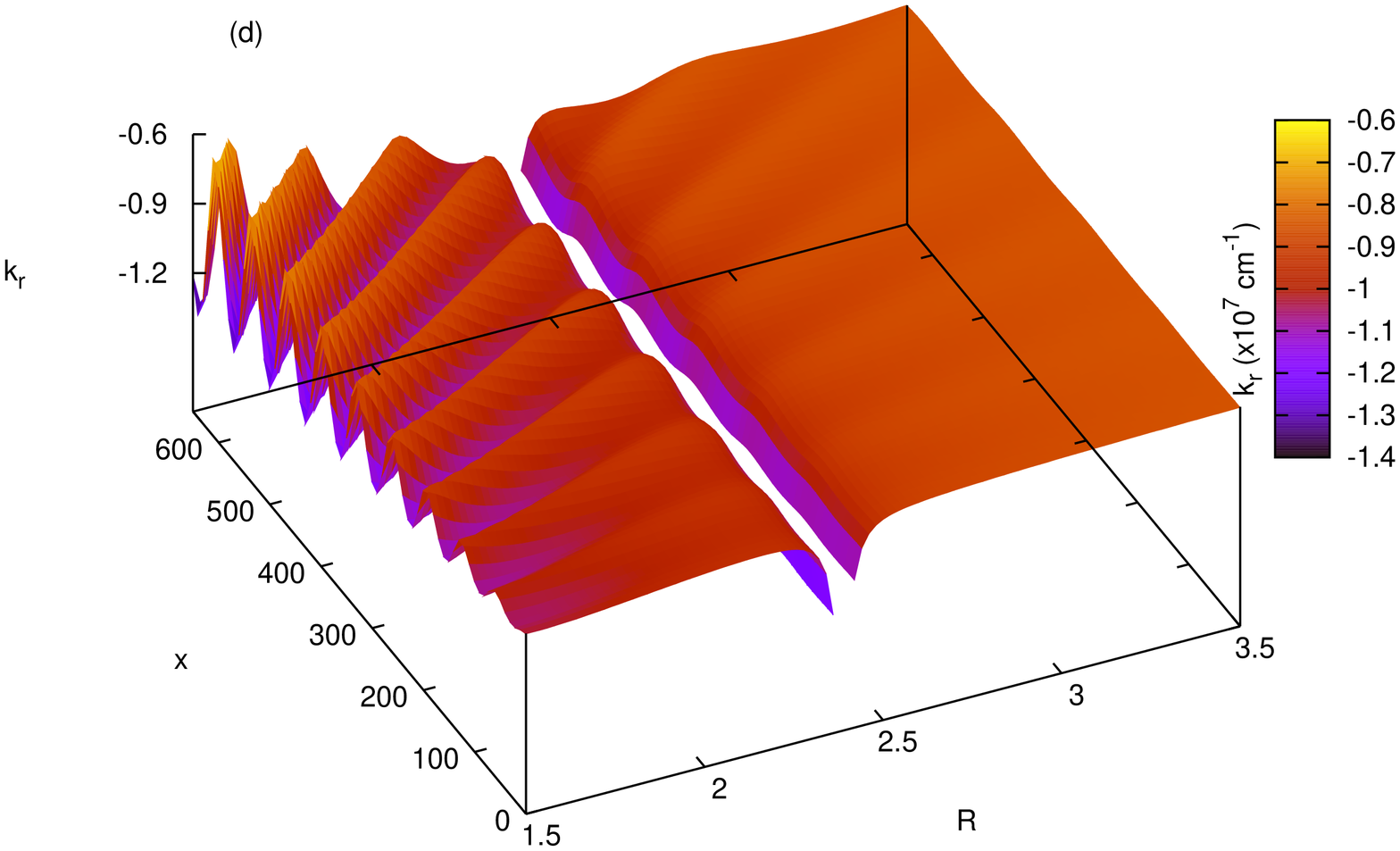}} \\
    \end{tabular}
    \caption{Two roots of the second order dispersion relation given by Eq. (24). (a) Forward propagating mode when the gradients of plasma parameters are not taken into account; (b) the same mode with gradients considered; (c) backward propagating mode without gradients and (d) the same mode with  gradients considered. The graphs are for ICW with a frequency 2500 $\rm rads^{-1}$. The higher values of $k_{r}$ in interplume lanes reveals higher refractive index $(n=ck_{r}/\omega)$ which in turn, points to the more effective wave-particle interaction and thus resonance.}
     \end{center}
\end{figure*}

From the solution of DR I we find that the wave numbers are of the
order of $10^{-8}-10^{-7} \rm cm^{-1}$ at the resonance site. By
putting this value into Eq. (21) we see that the highest value of
the residual contribution turns out to be $10^{-5}$. This is
negligibly small compared to the values of the rest of the terms
which range between $10^{6} - 10^{13}$ in the Eq. (22), so we can
neglect it. In this case, DR I given by the Eq. (23) is reduced to
DR II given by the Eq. (24) below,

\begin{figure}
\centering
\includegraphics[angle=270,scale=0.3]{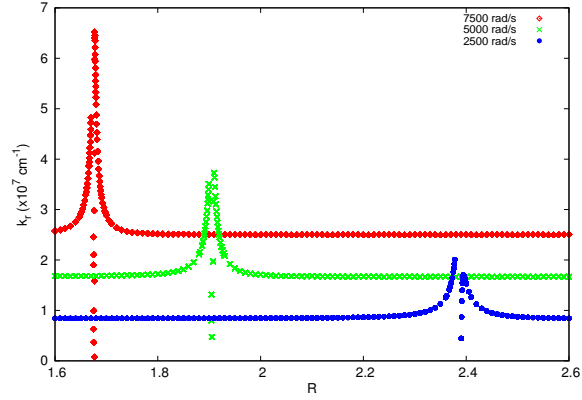}
\caption{This is the ($R, k_{r}$) version of Figure 3 (a). A local maximum is apparent close to the R = 2.4. At R = 2.38 the wave resonate with and transfer its energy to OVI ions.}
\end{figure}

\begin{equation}
 {{\it k}^{2} \left[{\it c}^{2} -\frac{{\it i}\omega _{p}^{2} \, }{2\alpha_{\parallel}^{2} \, \left(\omega -\omega _{c} \right)^{2} } { \Pi}\right]-\frac{{\it i}\omega _{p}^{2} \, {\it k}}{\sqrt{\pi } \, \alpha _{\parallel} \, \left(\omega -\omega _{c} \right)} {\Pi}-\frac{{\it i}\omega _{p}^{2} \, \omega }{\left(\omega -\omega _{c} \right)} -\omega ^{2} +}
 { \frac{\pi^{2}\omega {\it q}_{{\rm i}} \, L } {{\it E}_{{\rm x}} v_{A} }\nabla p_{1}^{\rm r} =0}
\end{equation}

The graphical solution of the Eq. (24) is given in Figure 4. The
solution of  DR II gives the values of k of about $10 ^{- 7} \,\rm
cm ^{- 1}$. With this \emph{k} value the residual contribution
becomes even more negligible. But we happened to show, \emph{a
posteriori}, that the residual contribution can safely be neglected.
Figure 4a is the solution of the Vlasov equation not including the
$-\nabla p$ force. The solution reveals an infinity in the
refractive index, corresponding to a resonance, throughout the
region considered at about 2.38 R. In this case, refractive index in
(R, x) domain has the same value, before and after the 2.38 R where
it becomes infinity. When we take the PIPL structure into account,
the infinity in the refractive index persists in the same 2.38 R
distance but the differences in the refractive index of the plumes
and interplume lanes are also revealed (Fig. 4b). Figure 4b shows
the crests and the troughs in the refractive index, the former
corresponds to the interplume lanes and the latter to the plumes. We
may argue that the refractive index of the interplume lanes is
readily going to infinity indicating that the resonance process in
the interplume lanes is more effective than in the plumes. This
result is confirmed by the observations showing that the source of
the fast solar wind is interplume lanes.

Figure 5 is the ($R, k_{r}$) version of Figure 4a. A local maximum
is apparent close to the R = 2.4. At R = 2.38 the wave resonate with
and transfer its energy to O VI ions. The pattern seen in Figure 4
is also visible in Figure 4(a,b,c,d). In the wave-particle
interaction context, one cannot consider a single wave frequency but
a band of frequencies. Therefore, we assumed the presence of ICW in
PCH generated with a frequency range of $2500 - 10000\,\rm
rads^{-1}$. The waves with frequencies higher than $2500\,\rm
rads^{-1}$ hit their magnetic beaches at smaller R and the ones with
frequencies lower than $2500\,\rm rads^{-1}$ at greater R distances
(see Fig.4).

We should mention, in passing, that when we adopt $10^{-3}$ for
$N_{OVI}/N_{p}$, which is the value Vocks (2002) used in his model
for the kinetics of ions in the solar corona, the ratio of the wave
numbers $k_{\bigtriangledown p}/k$ range from 0.2 to 16.8 in the
distance range 1.5-3.5 R, where $k_{\bigtriangledown p}$ is the wave
number of the waves propagating in the presence of gradients
considered, i.e. $\nabla T_{eff}^{\xi}(R,x)$,$\nabla n(R,x)$. For
$N_{OVI}/N_{p} = 1.52\times10^{-6}$ which is the value given by
Cranmer et al. (2008), $k_{\bigtriangledown p}/k$ take values
between 0.7 and 1.6. We plotted the figures by adopting the later
value, i.e. Cranmer et al.'s (2008). Despite the three orders of
magnitude difference in O VI abundances given in the above two
references, it is apparent that gradients both in x and R direction
of plasma parameters shortens the wavelength of the waves.

Finally, after having obtained wave numbers by solving the DRI and
DRII, we checked if the second condition for \emph{cold plasma}
approximation, i.e., $v_{\parallel}\ll \omega/k_{\parallel}$, holds
true. For ion thermal velocity we take the parallel component of the
bi-Maxwellian distribution function given in Eq. (18), that is,
$\alpha_{\parallel}^{-1}=(2k_{B}T_{\parallel}/m_{i})^{1/2}$, and for
the parallel phase velocity, $\omega/k_{\parallel}$, where $\omega$
is assigned a value in the range $2500 \,\rm rads^{-1}< \omega <
10000 \,\rm rads^{-1}$ and $k_{\parallel}$ are derived from DRI and
DRII. With these values, we obtained that
$v_{\parallel}/(\omega/k_{\parallel})\sim 10^{-3}$. This condition
is fulfilled also for the whole frequency range.

\section{Conclusions}
In the present investigation, the response of O VI ions to ICW in
the PCH is studied. When the wave frequency is equal to the ion
cyclotron frequency, i.e., $\omega = \omega _{c}$, LCP waves
transfer their energy to the O VI ions. This is true for all the
ions, but O VI ions are preferentially heated.

All the PCH plasma parameters, $N_{\rm e}$, $N_{\rm p}$, $N_{\rm
OVI}$, \emph{B}, $T_{\rm eff}$, $T_{\rm eff}^{\xi}$ obtained by
SoHO, are functions of the radial as well as the \emph{x}-direction.
This property of the medium makes the refractive index a function of
position. When the wave frequency approaches to the ion cyclotron
frequency ICR process takes place and the wave energy is transferred
to the O VI ions. This is revealed by the value of the refractive
index going, in a jumpy manner, to infinity. Energy transfer from
ICW to O VI ions is in such a way as to increase the perpendicular
velocity in their helical trajectory around the magnetic field. This
causes the violation of the first adiabatic invariant, the magnetic
moment, and increases the magnetic mirror force, consequently
accelerate OVI  ions in radial direction. Inclusion of $\nabla p$ in
DR I doesn't significantly alter the solution, but it is otherwise
for DR II. As it is apparent from the figures that refractive index
in the interplume lanes are about 2 times higher than the ones in
the plumes. This result of ours is confirmed by the observational
data that the seat of the fast solar wind is interplume lanes. For
instance, Prasad et al. (2011) by using high spatial and temporal
resolution of AIA/SDO, searched for the wave-like disturbances in
plumes and interplume lanes in PCH. They concluded that wave
propagation speeds are higher in the interplume lanes than those of
plumes.

We showed that for propagation parallel to the magnetic field in
PCH, infinities in $n^{2}_{\parallel}$ occur at ion cyclotron
frequency. Now, considering the ICW with a frequency $2500\,\rm
rads^{-1}$ propagating towards weakening magnetic field,
\emph{i.e.}, from the coronal base out to the extended corona in
radial direction, it is apparent from Figure 3a that wave frequency
approaches to the local cyclotron frequency. Dispersion relation
then yields larger $k_{\parallel}$ values with $k_{i}>0$ which
indicate the absorption of ICW through cyclotron damping. The
geometry of this process is referred to as \emph{magnetic beach}
(Stix, 1992).

In this investigation, we tried to show that ICR process can heat
one of the ion species, \emph{i.e.} OVI. Although we consider a
single ion species, this process seems to hold true for other ion
species as well.

\section*{Acknowledgments}

The authors acknowledge and thank an anonymous referee, E.Devlen,
K.Yakut and D.Cole for their useful suggestions and comments. Our
special thanks go to copyright holder of Figure 1 in the present
study, i.e., Wiley-VCH GmbH \& Co. KGaA and the authors of the paper
(E.Devlen \& E.R.Pek\"{u}nl\"{u}, 2010, Astron Nachr., 331, 716).
Figure 1 is reproduced with their kind permission. SD appreciates
the support by the Turkish Academy of Sciences (T\"UBA) Doctoral
Fellowship. This study is a part of PhD project of SD.

\bibliographystyle{model2-names}






\end{document}